\documentclass[conference]{IEEEtran}
\IEEEoverridecommandlockouts
\usepackage{cite}
\usepackage{amsmath,amssymb,amsfonts}
\usepackage{algorithmic}
\usepackage{graphicx}
\usepackage{textcomp}
\usepackage{xcolor}
\usepackage{subfigure}
\usepackage[]{collab}
\usepackage{hyperref}

\collabAuthor{sw}{purple}{Shiwen Shan}

\def\BibTeX{{\rm B\kern-.05em{\sc i\kern-.025em b}\kern-.08em
    T\kern-.1667em\lower.7ex\hbox{E}\kern-.125emX}}
\begin{document}

\title{eWAPA: An eBPF-based WASI Performance Analysis Framework for WebAssembly Runtimes
}

\author{Chenxi Mao,~Yuxin Su\thanks{Yuxin Su is the corresponding author.},~Shiwen Shan and Dan Li\\
\IEEEauthorblockA{\textit{School of Software Engineering} \\
\textit{Sun Yat-sen University}\\
Zhuhai, China}
}

\maketitle

\begin{abstract}
WebAssembly (Wasm) is a low-level bytecode format that can run in modern browsers. With the development of standalone runtimes and the improvement of the WebAssembly System Interface (WASI), Wasm has further provided a more complete sandboxed runtime experience for server-side applications, effectively expanding its application scenarios. 
However, the implementation of WASI varies across different runtimes, and suboptimal interface implementations can lead to performance degradation during interactions between the runtime and the operating system. 
Existing research mainly focuses on overall performance evaluation of runtimes, while studies on WASI implementations are relatively scarce. 
To tackle this problem, we propose an eBPF-based WASI performance analysis framework. 
It collects key performance metrics of the runtime under different I/O load conditions, such as total execution time, startup time, WASI execution time, and syscall time. 
We can comprehensively analyze the performance of the runtime's I/O interactions with the operating system. 
Additionally, we provide a detailed analysis of the causes behind two specific WASI performance anomalies. 
These analytical results will guide the optimization of standalone runtimes and WASI implementations, enhancing their efficiency.
\end{abstract}


\section{INTRODUCTION}
In recent years, the maturation and development of browser platforms have given rise to increasingly complex and large browser applications, such as large 3D games, audio and video software, and online deep learning real-time training platforms, highlighting the importance of efficiently executing code in browsers. 
However, JavaScript (JS), as the only built-in language on browsers, does not adequately meet the efficiency requirements\cite{haas2017bringing}. 
Even though JS engines like Chrome's V8 and SpiderMonkey have adopted Just-In-Time (JIT) compilation techniques to enhance the execution efficiency of JS code, their optimization iterations still cannot keep up with the rapid growth of modern browser applications. 

To address the aforementioned issues, WebAssembly (Wasm) emerged \cite{haas2017bringing}. Designed as a low-level bytecode format for application compilation targets, Wasm was initially intended to run code on web applications close to native performance, enabling the deployment of high-performance applications written in compiled languages like C, C++, Rust, etc., onto the web. 
In 2019, Wasm was ratified by the World Wide Web Consortium (W3C) to become the fourth browser language after HTML, CSS, and JS \cite{w3c_wasm_pr_cn}.

In reality, Wasm is an abstraction of modern hardware, making it independent of language, hardware, and platform, while providing guarantees for type and memory safety. 
The emergence of standalone Wasm runtimes has pushed Wasm beyond the confines of browsers, offering a fast, scalable, secure, and sandboxed method for running the same code on various non-web environments \cite{spies2021evaluation}, such as serverless computing \cite{FaasmLightweight20,2023webassembly,kjorveziroski2023webassembly}, edge computing \cite{Cloud-EdgeContinuum2022,Nomad2021}, and smart contracts \cite{chen2022wasai}. 
Particularly, the development of the Wasm System Interface (WASI) standardization efforts \cite{wasi2019} has provided a more complete sandboxed running experience on the server side, effectively expanding the usage scenarios of Wasm runtimes.


However, existing standalone Wasm runtimes are still in their early stages and are more likely to exhibit performance issues compared to browsers \cite{jiang2023revealing}, which can be hard to detect during the testing phase and may adversely affect applications. 
When conducting extensive system-level I/O operations through WASI in runtimes, the implementation of the WASI interface varies across different runtimes, including aspects such as memory management, syscall types, and security checks. 
Inefficient interface implementations can impact the overall performance of applications during interactions between the runtime and the system, undermining Wasm's performance advantages. 
Current research focuses on the overall performance of Wasm test cases on standalone runtimes and on revealing inconsistent behaviors between runtimes, yet there is a lack of studies on runtime WASI implementation efficiency issues in massive I/O interaction scenarios. 

In this paper, we aim to propose an automated WASI I/O performance testing method to reveal performance issues during WASI's I/O interactions. 
From the perspective of WASI implementation mechanisms, we aim to provide robust technical support for enhancing the performance of standalone runtimes executing Wasm modules, inspiring future work on optimizing more efficient standalone runtimes and WASI implementations.
The main contributions of this paper are as follows:

\begin{itemize}
\item We develop a non-intrusive eBPF-based monitoring tool to profile WASI-enabled Wasm runtimes, which collects performance data from both Native and runtime during the I/O process.
\item We identify and analyze the abnormal performance data of WASI from multiple dimensions, by studying the performance differences between various runtimes and then pinpointing specific abnormal behaviors.
\item Furthermore, we conduct an in-depth root cause analysis of two specific anomalies identified within the runtime, providing actionable insights for optimizing WASI implementations.
\end{itemize}

\section{BACKGROUND}

\subsection{Wasm Core Design}

\subsubsection{Stack Virtual Machine Design}
The architecture of Wasm and its compiler implementations leverage key characteristics of stack-based virtual machines, including structured control flow and stack containers, to achieve compact program representations \cite{haas2017bringing}.
The function code in Wasm modules consists of a series of instructions that operate on an implicit operand stack, popping parameter values and pushing result values. Although the size of bytecode for register machines is only 26\% larger than that for corresponding stack machines \cite{VirtualMachine2008}, the stack machine approach is not necessarily faster, but it offers smaller bytecode, which is an extremely important design goal for internet-based operations \cite{WebAssemblyDesignFreeCodeCamp}.

The stack machine design also makes the code verification process simpler and more efficient. Before loading and compiling Wasm modules, browsers first check whether the code complies with Wasm standards and security features, such as whether it accesses illegal memory ranges and whether the return value types of functions are correct. By examining the actual data type of the top element of the stack after a function execution, the compiler can directly determine whether the identified return value data type of the function is correct.

\subsubsection{Linear Memory Design}
The main memory of Wasm is linear memory, resembling a large byte array, which is a contiguous block of bytes. 
It mimics the traditional computer memory model, allowing data to be accessed according to byte addresses, and can be read from and modified by both Wasm and JS\cite{musch2019new}. 
When instantiating a Wasm module at runtime, a memory instance is created, which is essentially an expandable JS ArrayBuffer \cite{yan2021understanding}, enabling the use and emulation of dynamic memory allocation \cite{wang2021empowering}. 
The design of the ArrayBuffer provides a boundary, limiting the memory that Wasm modules can directly access \cite{MozillaWasmMemory}. 
Each Wasm module is permitted to define a single memory space, which may be shared across multiple instances.
As Wasm modules run, shared memory may experience memory overflow due to insufficient remaining resources, and Wasm allows for dynamic adjustment of the size of this shared memory.

\subsubsection{Sandbox Design}
When Linear memory is isolated, it does not share or overlap with other internal data structures of the execution engine, execution stacks, local variables, or memory of other processes \cite{WebAssemblyLinearMemory}, ensuring the security and isolation of memory operations and facilitating a sandboxed design. 
Untrusted modules can be securely executed within the same address space as other code, thereby facilitating rapid process-level isolation \cite{haas2017bringing}. 
This also allows Wasm to efficiently interact with untrusted JS and Web APIs and be embedded into other runtimes without compromising memory safety.

\subsection{Server-side Wasm}

With the growing popularity of server-side Wasm applications and the development of the WASI interface, the Wasm development community, such as the Bytecode Alliance \cite{BytecodeAlliance}, has implemented many standalone Wasm runtimes. 
Currently, there are over thirty independent Wasm runtimes available on GitHub \cite{jiang2023revealing}. 
For example, WebAssembly Micro Runtime (WAMR) \cite{WAMRDocumentation} is a lightweight standalone Wasm runtime known for its small footprint, high performance, and configurability. Wasmer \cite{WasmerGitHub} is an extremely fast and secure runtime that enables lightweight containers to run on desktops, clouds, edge computing, and even in browsers. Wasm3 \cite{Wasm3GitHub}, as a fast Wasm interpreter and the most universal Wasm runtime, uses a slow interpreter instead of fast just-in-time compilation. WasmEdge \cite{wasmedge} brings cloud-native and serverless application paradigms to edge computing. Wasmtime \cite{wasmtime} is a fast and secure Wasm runtime developed by the Bytecode Alliance.
All of these share the following fundamental components:

\subsubsection{Wasm Module Compilation Process}
The WASI-SDK \cite{WASISDKGitHub} is a toolkit developed for WASI, including specific compilers and libraries that optimize the compilation process. This SDK is used to compile C/C++ code into Wasm modules that conform to the WASI specifications. Wasm modules compiled with the WASI-SDK can access operating system-level functions, greatly expanding Wasm's application potential in server-side and other non-browser environments.

\subsubsection{WASI Interaction with Operating Systems}
Wasm is an assembly language intended for conceptual machines rather than physical ones, requiring a system interface of a conceptual operating system, allowing it to run on all different operating systems. 
With the help of WASI, Wasm applications can operate independently, securely communicating with the system without needing to be embedded in other programming environments \cite{spies2021evaluation}. 
Serving as a modular system interface between Wasm modules and the host operating system \cite{wasm_abis_guide}, WASI enables syscalls in a permission-based manner \cite{wasm_abis_guide}, contributing a series of implementation proposals for interactions in I/O, file systems, HTTP protocols, command-line interfaces, and network connections \cite{wasi_proposals}\cite{WebAssemblySpecs}.

\section{METHODOLOGY}\label{imple}
\subsection{Overview}
\begin{figure}[]
  \centering
  \includegraphics[width=0.5\textwidth]{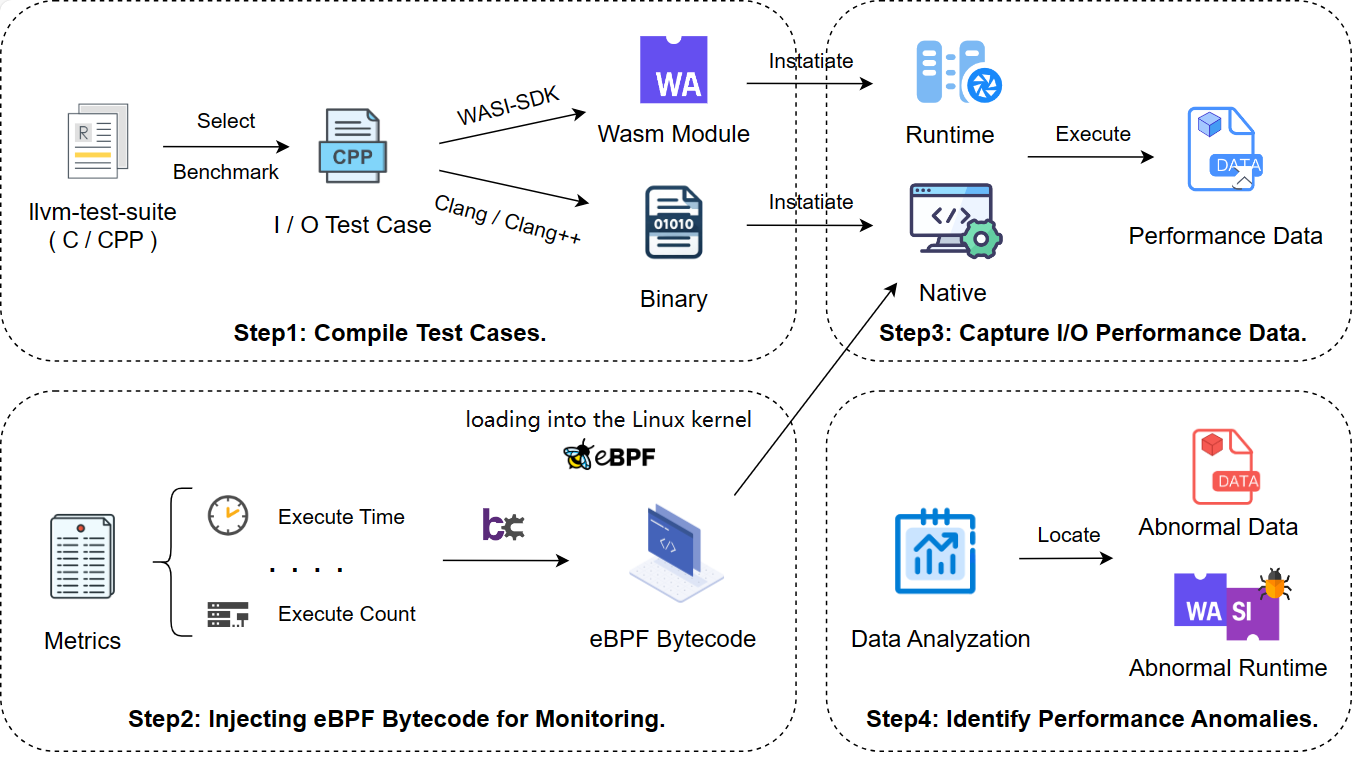}
  \caption{Method Overview.}
  \label{fig:overview}
\end{figure}

We develop an I/O Performance analysis method for Wasm runtimes based on eBPF, which identifies performance issues in the WASI implementation within standalone Wasm runtimes by running I/O-related test cases at runtime. 
This method, as shown in Figure \ref{fig:overview}, is divided into four key parts: compiling test cases, injecting eBPF bytecode into the Linux kernel to monitor performance, acquiring I/O performance data, and identifying anomalies in performance implementation.

Step 1 selects I/O test cases from the LLVM test suite, then compiling them into Wasm modules and native binary files. Step 2 writes eBPF programs and injects them into the Linux kernel to monitor performance metrics such as execution time and frequency of WASI and syscalls, as well as runtime startup time. Step 3 selects suitable test runtimes to execute the test cases on, and collects I/O performance data. Step 4 analyzes the data and determines the causes of performance anomalies in the runtime WASI.

\subsection{Step 1: Compile Test Cases.}
The first step is to compile test cases, which is divided into two sub-steps.

Initially, it is necessary to identify I/O test cases. To select test cases that well support Wasm and are more likely to trigger performance issues \cite{jiang2023revealing}, we utilize the LLVM test suite, which is designed for testing and evaluating the performance of LLVM compilers and toolchains \cite{llvmtestsuiteguide}. Test cases are chosen from the SingleSource directory of the test suite, as the source programs in this directory are individual programs written in C/C++, facilitating their compilation into Wasm and execution during the experiment. The test suite's cases primarily target terminal I/O; however, we focus on file I/O. Therefore, we modify the \textit{stdin} and \textit{stdout} in the test cases to use \textit{fopen} and \textit{fclose}, retains \textit{fwrite} and \textit{fread} functions, and add the execution of \textit{fseek} function to accommodate the characteristics of file I/O. The two test cases ultimately selected meet the requirements for testing various basic functions related to I/O.

Secondly, it is necessary to obtain the Wasm modules of the test cases and the control group's native binary files. For the Wasm runtime, we use WASI-SDK-21.0 to compile the source programs into Wasm modules. The WASI-SDK is a software development kit specifically developed for the WASI environment, providing a set of tools and libraries that enable developers to compile C and C++ code into Wasm, while also supporting system-level calls through the WASI API. For the native environment, we employ Clang and Clang++ tools to compile the source programs into native binary files with an optimization level of O3.

\subsection{Step 2: Injecting eBPF Bytecode into the Linux Kernel to monitor Performance.}
We has established a series of metrics, including WASI execution time, syscall execution time, total runtime, number of WASI executions, number of syscalls, average execution time of WASI, average execution time of syscalls, and runtime startup time. An eBPF program is written to monitor these metrics, typically using the C language.

The process of injecting eBPF into the kernel is commonly referred to as "loading." Specifically, this process involves using the BCC framework to compile the BPF program into bytecode, which is then loaded into the Linux kernel via a BPF syscall and attached to a kernel hook point, such as a network interface, syscall, or other kernel events.

In this paper, the primary focus of the hooks encompasses the runtime’s WASI function compilation symbols along with Linux’s I/O syscalls and symbols related to the compilation of runtime startup functions.
The WASI function and runtime startup-related function compilation symbols are hooks in user space, while I/O syscalls are hooks in the Linux kernel space. 
The eBPF monitoring symbols related to WASI in both native and Wasm runtimes are shown in Table \ref{tab:fuhao} (using the Read event as an example).

\begin{table*}[!htb]
    \caption{WASI FUNCTION COMPILATION SYMBOL HOOKS}
    \label{tab:fuhao}
    \centering
    \resizebox{1\linewidth}{!}{%
        \begin{tabular}{c|ccccccc|ccccccc}
        \hline
            & Native & Wasm3 & Wasmtime & wasmtime\_preview2 & Wasmer & WAMR \\ 
        \hline
        WASI Hook&fread&m3\_wasi\_generic\_fd\_read&wasi\_common...fd\_read& wasmtime\_wasi::preview2...fd\_read&wasmer\_wasix...fd\_read&wasi\_fd\_read\\ 
        Syscall Type&read&readv&readv&readv&read&pread64\\ 
        Syscall Hook&\_\_x64\_sys\_read&\_\_x64\_sys\_readv&\_\_x64\_sys\_readv&\_\_x64\_sys\_readv&\_\_x64\_sys\_read&\_\_x64\_sys\_pread64\\ 
        \hline
        \end{tabular}
    }
\end{table*}

\begin{table*}[!htb]
    \caption{WASM RUNTIME STARTUP-RELATED FUNCTIONS}
    \label{tab:init}
    \centering
    \resizebox{1\linewidth}{!}{%
        \begin{tabular}{c|ccccccc|ccccccc}
        \hline
            & Wasm3 & Wasmtime \& Preview2 & Wasmer & WAMR \\ 
        \hline
        Initialization Function&m3\_NewEnv...&new.execute()&\_\_libc\_start\_main&wasm\_runtime\_full\_init\\ 
        Function for Loading Wasm&repl\_load(...)&self.run.load\_module&WaAm::Module::imports&bh\_read\_file\_to\_buffer\\ 
        \hline
        \end{tabular}
    }
\end{table*}

To monitor the Read WASI execution time in the Wasmer runtime, first, locate the compiled symbol of the Read WASI function in the runtime flame graph, and use the \textit{nm} command to find the corresponding symbol in the runtime executable. 
In a Python program under the BCC framework, we use \textit{attach\_uprobe} to attach the eBPF program to user space. 
When the symbol's entry and exit points are hit, the corresponding \textit{trace\_fread\_entry} and \textit{trace\_fread\_exit} eBPF programs execute, storing data such as execution time and counts using the BPF\_Hash structure. 
Finally, data is retrieved and processed in Python code from the eBPF.

For the Read syscall, use the \textit{attach\_kprobe} function to attach to Linux kernel space, executing \textit{syscall.attach\_kprobe(event="\_\_x64\_sys\_read", fn\_name="t-\\race\_readv\_entry")} to attach the eBPF program to the kernel space symbol \textit{\_\_x64\_sys\_read}, filtering the triggering process's command to select the corresponding syscalls executed by the Wasm runtime.

Additionally, the eBPF monitoring functions related to Wasm runtime startup time are as shown in Table \ref{tab:init}. When the runtime begins loading the Wasm module, it indicates that the runtime has completed initialization. The startup time is the interval from the entry of the initialization function to the entry of the function loading the Wasm module. Taking the Wasm3 runtime as an example, first, initialize result variables and the environment, execute \textit{env = m3\_NewEnvironment();} to create a new Wasm3 environment, then parse command line arguments, using a series of conditional statements to identify different command flags and set the corresponding variables. Next, perform module loading and initialization. The program, based on parsed parameters, calls \textit{repl\_init}to initialize the REPL (Read-Eval-Print Loop) environment, set the stack size and then uses \textit{repl\_load} to load the specified Wasm module. Finally, perform conditional compilation, module execution, and error handling.

\subsection{Step 3: Acquiring I/O performance data.}


We execute the Wasm modules compiled in Step 1 on various test runtimes. 
At the same time, the compiled native binary files are run in the local environment as a control. 
The evaluation of the fread function, alongside the overall runtime, is conducted using three distinct input file sizes: 1GB, 10GB, and 100GB. Similarly, the fwrite function and the total runtime are assessed across four varying output file sizes: 48MB, 4.7GB, 11GB, and 99GB.
Additionally, the startup time of the runtime is tested under two different input file sizes of 1GB and 10GB, and two different output file sizes of 48MB and 4.7GB. To prevent auxiliary I/O functions from exhibiting performance issues due to too short startup times, we execute the \textit{fseek}, \textit{fopen}, and \textit{fclose} functions 50 times each in a loop to better identify performance issues. For \textit{fseek}, the test file size is 1GB, with the loop count set as \(i\), so the seek position each cycle is \(i \times 1,000,000\); for \textit{fopen} and \textit{fclose}, the test file size is 4.7GB. In each runtime and native environment, each test case Wasm module is executed 10 times, and the average of the 10 execution times is taken as the final execution time.

\subsection{Step 4: Identifying performance anomalies.}
This step analyzes the I/O performance data collected in Step 3, identifying anomalies and the causes of performance issues in runtime WASI. It is divided into four substeps:

\textbf{(1) Analyzing runtime execution time trends under varying I/O loads.} We collect data on I/O syscall times, I/O WASI interaction times, runtime startup times, and total runtime execution times at different I/O file sizes. Line charts depicting the trend of increasing execution times as file sizes increase are drawn, and the charts are analyzed to draw some conclusions.

\textbf{(2) Analyzing the proportion of runtime execution time under heavy I/O loads.} We calculate the proportion of \textit{Read/Write} WASI and syscalls in the total runtime for inputs of 100GB and outputs of 99GB and chart these proportions. Additionally, to compare the impact of different I/O loads on these proportions, we also calculate and chart the proportions of \textit{Read/Write} WASI and syscalls in the total runtime for inputs of 10GB and outputs of 4.7 GB.

\textbf{(3) Analyzing the WASI execution time for I/O auxiliary functions.} The WASI execution times for the fopen, close, and fseek functions are analyzed.

\textbf{(4) Case study.} Data related to I/O WASI and syscalls for different runtimes under large file I/O conditions (input 100GB, output 99GB) are collected. This includes a series of metrics such as WASI's execution time, number of executions, average execution time, and syscall's execution time, number of executions, and average execution time. We select two instances of performance anomalies from the experimental results and find out the reasons by analyzing data and code implementation.

\section{EVALUATION AND ANALYSIS}
In this section, we aim to answer the following research questions:

\begin{itemize}
    \item RQ1: How does the trend of runtime execution times vary with different I/O loads?
    \item RQ2: When the I/O load is large, what is the proportion of runtime execution time?
    \item RQ3: What is the execution time of WASI for I/O auxiliary functions?
\end{itemize}
\subsection{Experimental Setting}
\textbf{Experimental Environment.} All experiments were conducted on a server equipped with an Intel(R) Core(TM) i5-12400F 4.4GHz CPU and 32GB of DDR4 memory. The server operates on a 64-bit Arch Linux system, with the Linux kernel version 6.8.2-arch2-1. 

\textbf{Wasm Runtime Selection.} We search for the keyword “WebAssembly Runtime” on GitHub, and select runtimes with stars more than 4.5K. We also filter out runtimes without continuous releases and acceptable execution time.
Ultimately, four representative independent Wasm runtimes are selected: Wasmer, Wasmtime, Wasm3, and WAMR. During the experiment, Wasmtime updated to the preview2\footnote{The wasmtime\_wasi package is transitioning from the older wasi\_common (preview1) to supporting wasmtime\_wasi::preview2. This new implementation supports both WASIp1 and WASIp2 versions. This transition signifies a gradual deprecation of the old implementation, with a full shift towards a more modern and updated WASI implementation in the future.} version \cite{wasmtimev18}, we consider it as a fifth runtime. Information on the runtimes is shown in Table \ref{tab:wasmruntime}.

\begin{table}[!htb]
    \caption{WASM RUNTIME VERSION INFORMATION}
    \label{tab:wasmruntime}
    \centering
    \resizebox{1\linewidth}{!}{%
        \begin{tabular}{c|ccccccc|ccccccc}
        \hline
            & GitHub Stars & Test Version & Commit Hash \\ 
        \hline
        Wasm3 & 7k & - & 772f8f4648fcba75f77f894a6050db121e7651a2  \\ 
        WAMR & 4.5k & - & 52db362b897221c2a438197a0f2e4a9a300becd4  \\ 
        Wasmtime & 14.4k & v14.0.4 & -    \\ 
        Wasmer & 17.7k & v4.2.2 & -    \\ 
        Wasmtime\_preview2 & 14.4k & v18.0.0 & -    \\ 
        \hline
        \end{tabular}
    }
\end{table}


\subsection{RQ1: Execution Time Trend}
\subsubsection{Syscall Execution Times}\label{syscall}
\begin{figure}[htbp]
\begin{center}
\subfigure[Fread syscall time trend]{
\includegraphics[width=0.46\linewidth]{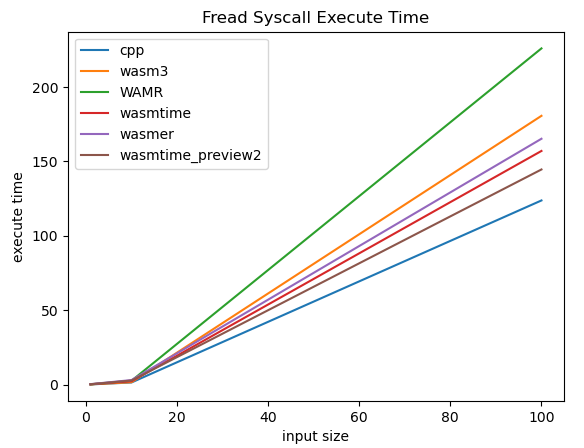}
}
\subfigure[Fwrite syscall time trend]{
\includegraphics[width=0.46\linewidth]{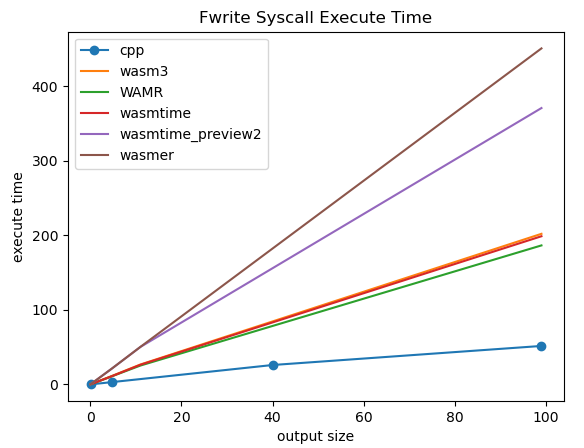}
}
\end{center}
\caption{\textbf{I/O syscall time trend (seconds)}}
\label{fig:rq1syscall}
\end{figure}

With different I/O loads, the trends in \textit{read} and \textit{write} syscall times for native and Wasm runtimes are illustrated in Figure \ref{fig:rq1syscall}.

\textcircled{1} Compared to \textit{write}, the syscall times for \textit{read} do not show a completely linear growth trend as sizes of input file increase. The rate of increase in execution time of syscall for reading large files is greater than that for smaller files in both native environments and various runtimes. 
This suggests that runtimes are more likely to encounter bottlenecks when reading large-scale data. 
Conversely, the syscall times for \textit{write} exhibit a more stable growth trend with increasing file sizes.

\textcircled{2} In read operations, the syscall times across the four runtimes are generally similar, though WAMR's syscall time is noticeably slightly higher than the others. In \textit{write}, the syscall times for Wasmtime, WAMR, and Wasm3 are quite close, while Wasmer and Wasmtime\_preview2 show significantly higher times compared to the first three.

\textcircled{3} There are some commonalities in both \textit{read} and \textit{write}: the syscall times for native \textit{read} and \textit{write} are the shortest among all runtimes, which aligns with normal expectations.

\subsubsection{WASI Execution Times}\label{wasi} 
\begin{figure}[htbp]
\begin{center}
\subfigure[Fread WASI time trend]{
\includegraphics[width=0.46\linewidth]{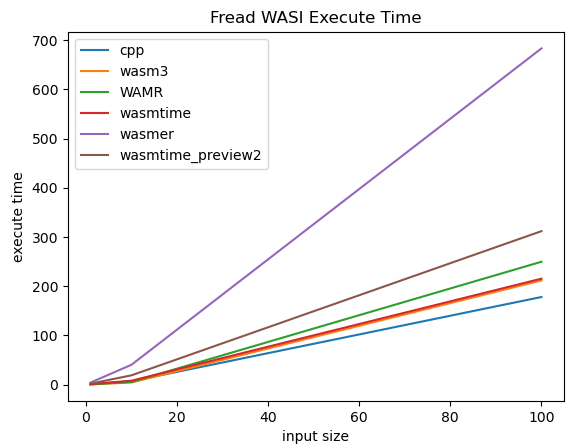}
}
\subfigure[Fwrite WASI time trend]{
\includegraphics[width=0.46\linewidth]{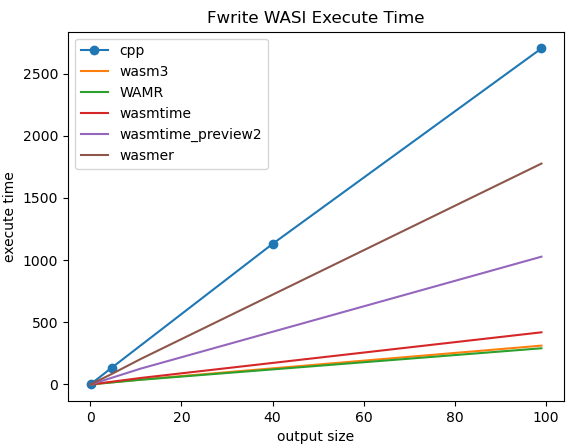}
}
\end{center}
\caption{\textbf{I/O WASI time trend (seconds)}}
\label{fig:rq1wasi}
\end{figure}

Under different I/O loads, the trends in WASI interaction time during \textit{read} and \textit{write} for Native and Wasm runtimes are illustrated in Figure \ref{fig:rq1wasi}. The horizontal axis represents the unit of GB, and the vertical axis represents the unit of seconds. Regarding the analysis of WASI execution time, significant performance differences in WASI and system interaction are observed across various runtimes when handling different scales of file \textit{read} and \textit{write}.


\textcircled{1} When Native performs file \textit{write}, the \textit{fwrite} function exhibits exceptional performance behavior, showing the highest execution time growth rate. 
As shown in Section \ref{syscall}, the syscall time for Native is the shortest. This indicates that in the Native environment, there exists much redundant time between syscalls and the \textit{fwrite} function, resulting in longer execution time for \textit{fwrite} compared to WASI. Additionally, when handling small-scale file \textit{read}, although the execution time of Native's \textit{fread} function is longer, slightly higher than Wasm3 and WAMR, it does not show a significant performance advantage. 
However, for larger input sizes, Native's \textit{fread} function typically demonstrates the lowest execution time. 
This suggests that compared to other runtimes, Native's \textit{fread} function may have better input processing efficiency for large files, while struggling to showcase performance advantages for small files.

\textcircled{2} Compared to \textit{write}, the WASI time for \textit{read} does not exhibit a completely linear growth trend with increasing input file sizes. The growth rate of time for large files of each runtime is higher than that for small files, indicating potential bottlenecks when processing larger-scale data, consistent with the first point in Section \ref{syscall}.

\textcircled{3} Some commonalities are observed in both \textit{read} and \textit{write}: Wasm3, Wasmtime, and WAMR exhibit similar execution time growth curves, with close WASI execution times. 
During file \textit{read}, Wasmer shows the highest execution time growth rate, suggesting that as the input size increases, its performance degradation in WASI and system interaction may be the most significant, followed by Wasmtime\_preview2. 
During file \textit{write}, Wasmer and Wasmtime\_preview2 still exhibit relatively poor interaction performance.

\subsubsection{Runtime Startup Time}
\begin{figure}[!htb]
    \centering
    \includegraphics[width=0.4\textwidth]{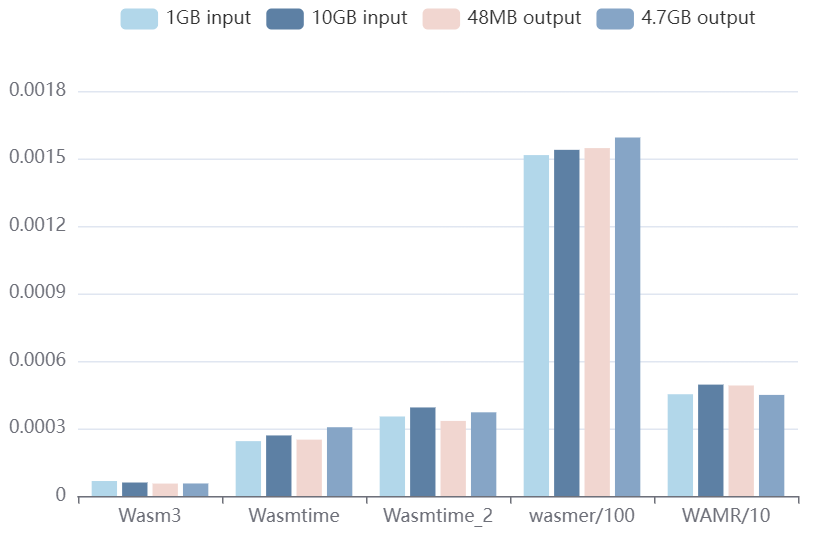}
    \caption{Wasm runtime startup time under different I/O scales (seconds)}
    \label{fig:init}
\end{figure}

Figure \ref{fig:init} illustrates the startup times of various runtimes. The horizontal axis represents the names of the runtimes, and the vertical axis shows the execution time in seconds. 
Due to the excessively long startup times of Wasmer and WAMR, the data for these two runtimes were adjusted by reducing them by factors of 100 and 10, respectively, to better display the data for other runtimes in the figure.

\textcircled{1} Comparing the startup times of the same runtime under different I/O scales reveals that even with input and output scales differing by more than tenfold, the variations in startup times for executing different test cases across various Wasm runtimes are not significant. 
This suggests that these runtimes have low dependencies on input and output scales during startup. 
It also indicates that these runtimes may be well-optimized during startup, with minimal increase in startup time due to data scale.

\textcircled{2} Contrasting the startup times of different runtimes under the same I/O scale shows significant differences. 
Wasm3 has the shortest startup time, followed by Wasmtime and Wasmtime\_preview2, while the startup times of WAMR and Wasmer are significantly higher than the other three runtimes. 
Wasmtime and Wasmtime\_preview2, as different implementations of the same runtime, exhibit similar startup times. 
Although Wasm3 and WAMR show similar performance on several other tests, the difference in startup time is substantial, nearly a hundredfold. Wasmer's startup time is also much longer than WAMR and the other three runtimes, differing by almost 2000 times from the shortest startup time of Wasm3.

The evaluation of runtime startup times indicates that the startup times remain nearly unchanged in response to changes in the I/O scale. It means that the startup phase is well-optimized and largely independent of data size. However, significant differences in startup times between different runtimes are apparent, with some runtimes exhibiting exceptionally long startup times.

\subsubsection{Overall Execution Time}\label{all}
\begin{figure}[htbp]
\begin{center}
\subfigure[Fread overall execution time trend]{
\includegraphics[width=0.46\linewidth]{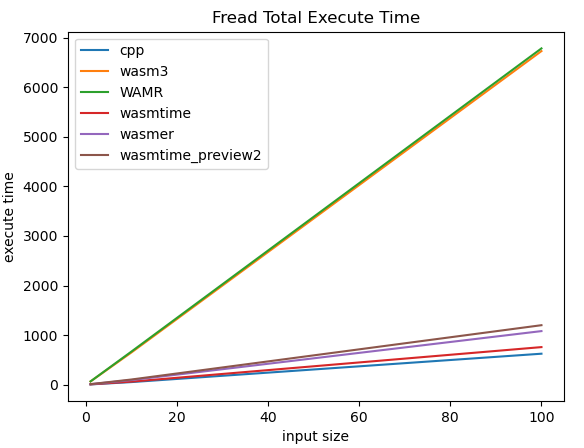}
}
\subfigure[Fwrite overall execution time trend]{
\includegraphics[width=0.46\linewidth]{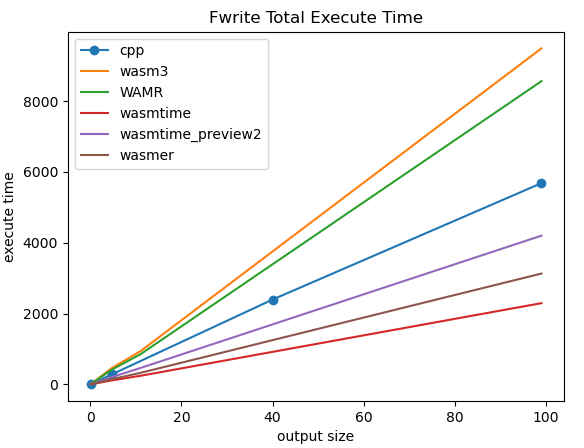}
}
\end{center}
\caption{\textbf{I/O overall execution time trend (seconds)}}
\label{fig:rq1all}
\end{figure}

Under different I/O loads, the overall execution time trends for \textit{read} and \textit{write} in Native and Wasm runtimes are illustrated in Figure \ref{fig:rq1all}. The horizontal axis represents GB, and the vertical axis represents seconds.


\textcircled{1} For both \textit{read} and \textit{write}, the execution times of the WAMR and Wasm3 are close. 
As the I/O scale increases, the overall runtime time shows a significant linear growth trend. 
This may indicate that both runtimes face performance bottlenecks when handling large amounts of data, possibly due to inefficient implementation of code outside the WASI.

\textcircled{2} During \textit{read}, apart from WAMR and Wasm3, other runtimes show a more moderate increase in execution time. 
The execution times of Wasmer and Wasmtime\_preview2 are close and slightly higher than Native's. 
Wasmtime’s execution time is the closest to Native, demonstrating faster running speed and overall efficiency. 
In \textit{write}, the execution times of Wasmer, Wasmtime\_preview2, and Wasmtime are shorter than the total execution time of Native. 
The reduction in total execution time for Wasmer and Wasmtime\_preview2 is partly due to the impact of the Tokio asynchronous mechanism used by these runtimes, which helps reduce overall execution time during large file I/O operations.

\subsection{RQ2: Proportion of Execution Time}\label{rq2}
\begin{figure}[!htb]
    \centering
    \includegraphics[width=0.4\textwidth]{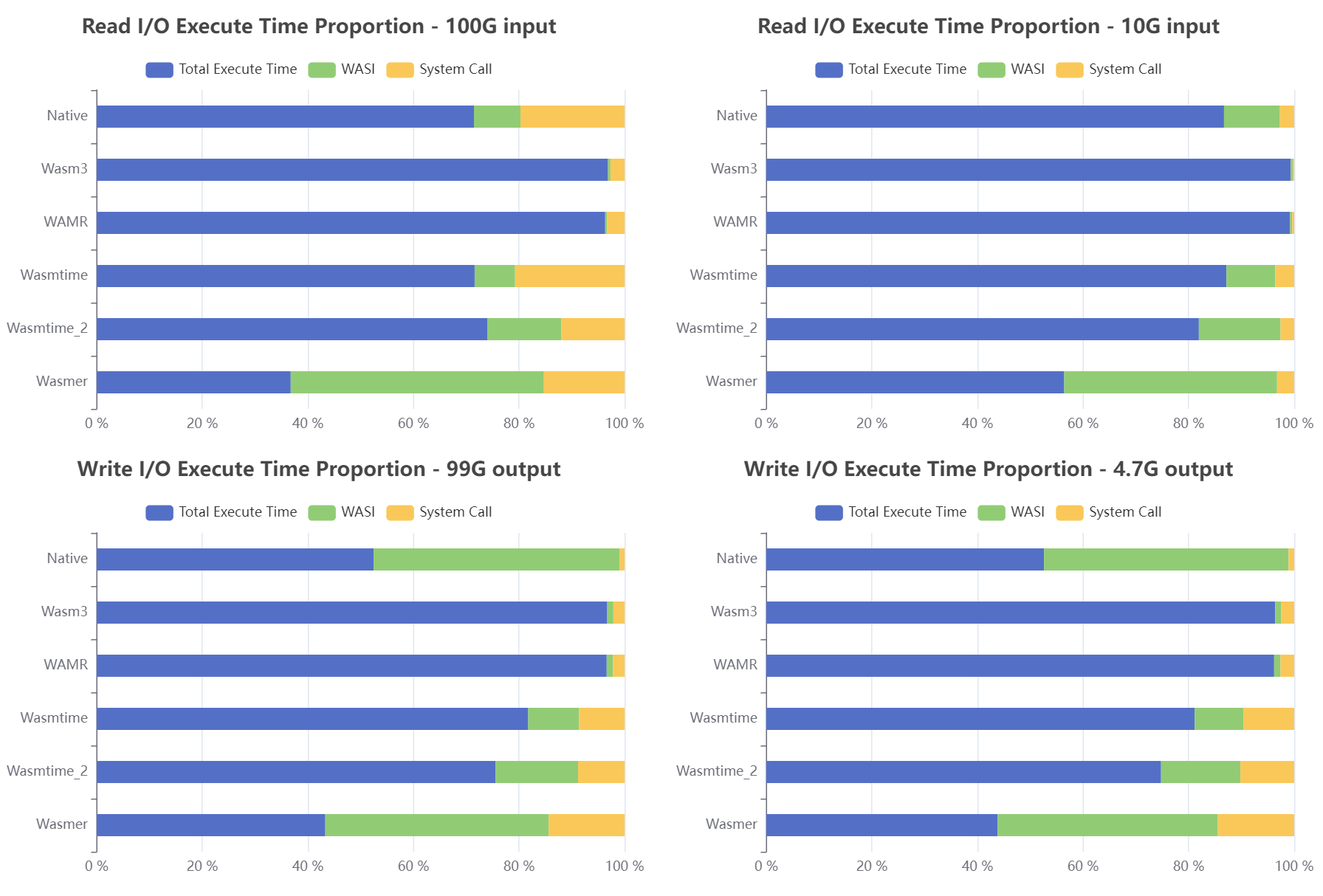}
    \caption{Distribution of proportion of time occupied by each part of I/O.}
    \label{fig:RQ2}
\end{figure}

We develop a frontend program to display the proportion of WASI and syscall times in the total runtime, as shown in Figure \ref{fig:RQ2}.
Comparing large-scale \textit{read} and \textit{write}, it can be seen that the runtime, excluding WASI, accounts for a high proportion in Wasm3 and WAMR, which aligns with the first point in Section \ref{all}.
The execution times of Native’s \textit{fwrite}, Wasmtime’s I/O WASI, Wasmtime\_preview2’s I/O WASI, and Wasmer’s I/O WASI have a high proportion in the overall runtime, especially in Wasmer.

When comparing the \textit{write} operation across different I/O scales, it is observed that the proportion of execution times remains relatively consistent across all evaluated runtimes.
An analysis of \textit{read} operations across various I/O scales indicates that, particularly for large-scale \textit{read} processes, there is a substantial increase in the proportion of WASI and syscall execution times relative to the total runtime.
This aligns with the second point in Section \ref{wasi}.

\subsection{RQ3: Execution Time of I/O Auxiliary Functions}\label{rq3}
We test lightweight I/O auxiliary functions such as \textit{fseek}, \textit{fopen} and \textit{fclose}, and obtain the interaction times with WASI for each function.

\subsubsection{Fopen and Fclose}
By repeatedly executing \textit{fopen} and \textit{fclose} for 50 times, we obtain the interaction times with WASI, as shown in Table \ref{tab:fopen}. 
For \textit{fopen}, the execution times of Native and most Wasm runtimes are generally consistent, with Wasmtime\_preview2 having significantly shorter execution times than the other runtimes. 
For \textit{fclose}, the execution times of Wasmtime\_preview2 and Wasmer are much shorter than those of the other runtimes, followed by Wasm3 and WAMR.

\begin{table}[!htb]
    \caption{WASI INTERACTION TIME BETWEEN FOPEN And FCLOSE (Seconds)}
    \label{tab:fopen}
    \centering
    \resizebox{1\linewidth}{!}{%
        \begin{tabular}{c|ccccccc|ccccccc}
        \hline
            & Native & Wasm3 & Wasmtime & Wasmtime\_preview2 & Wasmer & WAMR \\ 
        \hline
        Fopen & 20.86419 & 21.09183 & 19.69448   & 0.00036 & 21.45274 & 21.10804 \\ 
        Fclose & 35.00267 & 5.17304 & 24.15975 &0.00016 & 0.00019 & 5.87338\\ 
        \hline
        \end{tabular}
    }
\end{table}

\subsubsection{Fseek}
The test file size for \textit{fseek} is 1GB, with 50 iterations. 
Let the iteration count be $i$, and the seek position for each iteration is $i*1000000$. 
Tests are conducted on Native and various Wasm runtimes to obtain the interaction times with WASI, as shown in Table \ref{tab:fseek}. 
The \textit{fseek} times for Wasm3, Wasmer, and WAMR are relatively close and do not exhibit significant performance anomalies. 
The \textit{fseek} execution times for Wasmtime Preview1 and Preview2 are relatively longer.

\begin{table}[!htb]
    \caption{WASI INTERACTION TIME Of FSEEK (Seconds)}
    \label{tab:fseek}
    \centering
    \resizebox{1\linewidth}{!}{%
        \begin{tabular}{c|ccccccc|ccccccc}
        \hline
            & Native & Wasm3 & Wasmtime & Wasmtime\_preview2 & Wasmer & WAMR \\ 
        \hline
        Fseek & 0.00033 & 0.00013 & 0.00023   & 0.00022 & 0.00017 & 0.00015 \\ 
        \hline
        \end{tabular}
    }
\end{table}

\section{CASE STUDY}
To explain the reasons behind some of the above conclusions, we select two performance anomaly cases and analyze their root causes. 
First, under large file I/O conditions (input 100GB, output 99GB), data related to I/O WASI and syscalls are collected for different runtimes. 
It includes the WASI execution time, execution count, average execution time, syscall execution time, execution count, and average execution time, as shown in Table \ref{tab:fread} and Table \ref{tab:fwrite}.

\begin{table*}[!htb]
    \caption{FREAD RELATED WASI And SYSCALL DATA}
    \label{tab:fread}
    \centering
    \resizebox{1\linewidth}{!}{%
        \begin{tabular}{c|ccccccc|ccccccc}
        \hline
            & Fread WASI Time (s) & WASI Numer (s) & WASI Average Time (s) & Read syscall Time (s) & syscall Num (s) & syscall Average Time (s) \\ 
        \hline
        CPP & 178 & 26214401 & 7719 & 123 & 26214401 & 5643 \\ 
        Wasm3 & 211 & 26214401 & 9244 & 180 & 26214401 & 8100\\  
        WAMR & 249 & 26214401 & 9660 & 226 & 26214401 & 9701\\
        Wasmtime\_preview2 & 311 & 52428801 & 6528 & 144 & 52428801 & 3324\\ 
        Wasmer & 683 & 26214401 & 26796 & 165 & 52428801 & 3333\\ 
        Wasmtime & 215 & 26214401 & 8088 & 157 & 26214401 & 5858\\ 
        \hline
        \end{tabular}
    }
\end{table*}

\begin{table*}[!htb]
    \caption{FWRITE RELATED WASI And SYSCALL DATA}
    \label{tab:fwrite}
    \centering
    \resizebox{1\linewidth}{!}{%
        \begin{tabular}{c|ccccccc|ccccccc}
        \hline
            & Fwrite WASI Time (s) & WASI Numer (s) & WASI Average Time (s) & Write syscall Time (s) & syscall Num (s) & syscall Average Time (s) \\ 
        \hline
        CPP & 2701 & 1750000077 & 1616 & 51 & 25976563 & 2533 \\ 
        Wasm3 & 311 & 101797388 & 3302 & 201 & 101797388 & 2191\\  
        WAMR & 290 & 101797388 & 3181 & 186 & 101797388 & 2129\\
        Wasmtime\_preview2 & 1026 & 203594775 & 5505 & 370 & 203594775 & 2165\\ 
        Wasmer & 1775 & 101797388 & 17261 & 370 & 203594776 & 2193\\ 
        Wasmtime & 419 & 101797388 & 3681 & 198 & 101797388 & 1597\\ 
        \hline
        \end{tabular}
    }
\end{table*}

\subsection{Case Analysis 1}
The first case analyzed is the longer execution time of the \textit{fwrite} function in the Native environment compared to the Write WASI interaction time of all runtimes. 
This is a counter-intuitive performance anomaly.

\begin{figure}[!htb]
    \centering
    \includegraphics[width=0.45\textwidth]{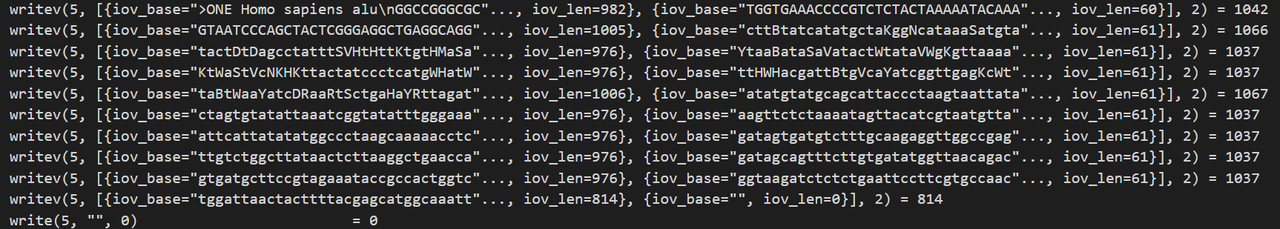}
    \caption{Wasm3 Syscall: writev}
    \label{fig:writev}
\end{figure}

According to Table \ref{tab:fwrite}, in terms of syscall time, the total syscall time for Native is much shorter than that for other runtimes. 
Native uses the \textit{write} syscall, with a total of 25,976,563 calls, fewer than the actual \textit{fwrite} calls of 1,750,000,077 times, with an average syscall time of 2533ns. 
In contrast, Wasm runtimes (e.g., wasm3) use the \textit{writev} syscall (as shown in Figure \ref{fig:writev}, with each \textit{writev} divided into two parts, writing approximately 976 bytes and 61 bytes, totaling 1037 bytes), with a total of 101,797,388 calls, consistent with the Write WASI call count, and an average syscall time of 2191ns. 
The difference in the writing methods of syscalls between Native and other runtimes results in a significantly shorter total syscall time for Native, despite the similar average syscall time per call. 
The discrepancy is due to differences in the mechanisms for writing cached data to disk between the two.


In the C/C++ standard library, the \textit{fwrite} function is typically equipped with an internal buffering mechanism. 
Data is first written to the internal buffer, and only when the buffer is full or an explicit flush is triggered (such as by calling the \textit{fflush} function), will the actual \textit{write} call to the operating system occur. 
This is also reflected in the timing of \textit{fwrite} executions: most \textit{fwrite} executions are very short, but occasionally, there will be longer execution times, which usually indicates an actual \textit{write}. 
When the buffer is full, after about 68 \textit{fwrite} calls, a single \textit{write} syscall writes the entire 4096 bytes to disk.

In Wasm, the buffer size is not fixed. 
Taking Wasm3 as an example (with similar behavior in other runtimes), according to the \textit{writev} syscall data shown in Figure \ref{fig:writev}, when the buffer accumulates to approximately 1037 bytes, the runtime will initiate a \textit{write} WASI call, which then triggers a \textit{writev} syscall to actually write the data to the file. 
Specifically, in the test case, each \textit{fwrite} writes 60 bytes, and 16 consecutive \textit{fwrite} calls accumulate to 976 bytes, corresponding to the first \textit{iov\_base} of the \textit{writev} syscall. 
An additional \textit{fwrite} call adds 61 bytes, corresponding to the second \textit{iov\_base}. Therefore, approximately 17 \textit{fwrite} function executions, totaling 1037 bytes, will trigger a \textit{write} WASI and a \textit{writev} syscall.

From Table \ref{tab:fwrite}, it can be seen that the number of \textit{fwrite} calls is 1,750,000,077 times, which matches the actual number of \textit{fwrite} function calls in the code. 
The number of \textit{write} WASI calls is 101,797,388 times, and the number of \textit{writev} syscalls is also 101,797,388 times. 
This essentially verifies the above ratio, i.e., the execution ratio of \textit{fwrite} to \textit{write} WASI to \textit{writev} is 17:1:1.

To verify the variability of the Wasm runtime buffer size, we modify the test case code to set each write to 280 bytes and re-traced the execution. 
In the test, each \textit{fwrite} writes 280 bytes, and three consecutive \textit{fwrite} calls accumulate 840 bytes, corresponding to the first \textit{iov\_base} of the \textit{writev} syscall. 
An additional \textit{fwrite} call adds 280 bytes, corresponding to the second \textit{iov\_base}. 
This indicates that approximately four \textit{fwrite} function executions, totaling 1120 bytes, will trigger a \textit{write} WASI and a \textit{writev} syscall. 
In this test case, the ratio of \textit{fwrite} to \textit{write} WASI to \textit{writev} will be 4:1:1. 
This result supports the view that the Wasm runtime buffer size is variable and different from the Native buffer size.


In conclusion, the significant differences in the execution time of \textit{fwrite} between Native and Wasm can be primarily attributed to the different buffering mechanisms. 
In the Native environment, the execution time of \textit{fwrite} includes the time taken to write data into a fixed 4096-byte buffer until it is full, followed by a single \textit{write} syscall. This means that many \textit{fwrite} that merely involves writing to the buffer is also included in the timing statistics. 
In contrast, in Wasm WASI, the buffer size is not fixed, and once the accumulated data reaches a certain \textit{iov\_base} size, a \textit{write} WASI execution and a \textit{writev} syscall are triggered. 
Statistically, Wasm WASI only records the execution times of those instances that actually involve file writes, which are closely related to the syscalls, more directly reflecting the actual disk \textit{write}. 
The ratio of \textit{fwrite} executions in Native to Wasm WASI is 17:1, so the \textit{fwrite} execution times recorded in the Wasm WASI environment typically appear shorter.

The differences in the design of buffering mechanisms between the C/C++ standard library and Wasm also explain why there is no performance discrepancy observed in the \textit{read} WASI tests. This is because the test case specifies that each \textit{read} is 4096 bytes, making the buffer size identical in both environments. 
Therefore, as shown in Table \ref{tab:fread}, with the number of \textit{read} being consistent, the execution time of \textit{fread} in the Native environment is shorter than that in Wasm WASI, as expected.



However, if the number of bytes read each time is adjusted to 2000 bytes, the situation changes. 
In the Native environment, each execution of the \textit{fread} function reads 2000 bytes from the buffer; when the buffer is empty (approximately after 2 \textit{fread} executions), the system performs a \textit{read} syscall to read another 4096 bytes from the disk to refill the buffer. This indicates that in the C/C++ standard library, the buffer for \textit{read} is also set to 4096 bytes. 
In the Wasm environment, each execution of the \textit{fread} function directly corresponds to a \textit{read} WASI call and a \textit{readv} syscall. 
Each \textit{readv} syscall reads 2000 bytes of file content. 
In this case, the \textit{fread} function in Native and the \textit{read} WASI in Wasm will, like the \textit{fwrite} function, result in differences in execution counts and recorded times.

\subsection{Case Analysis 2}
\begin{figure}[!htb]
    \centering
    \includegraphics[width=0.45\textwidth]{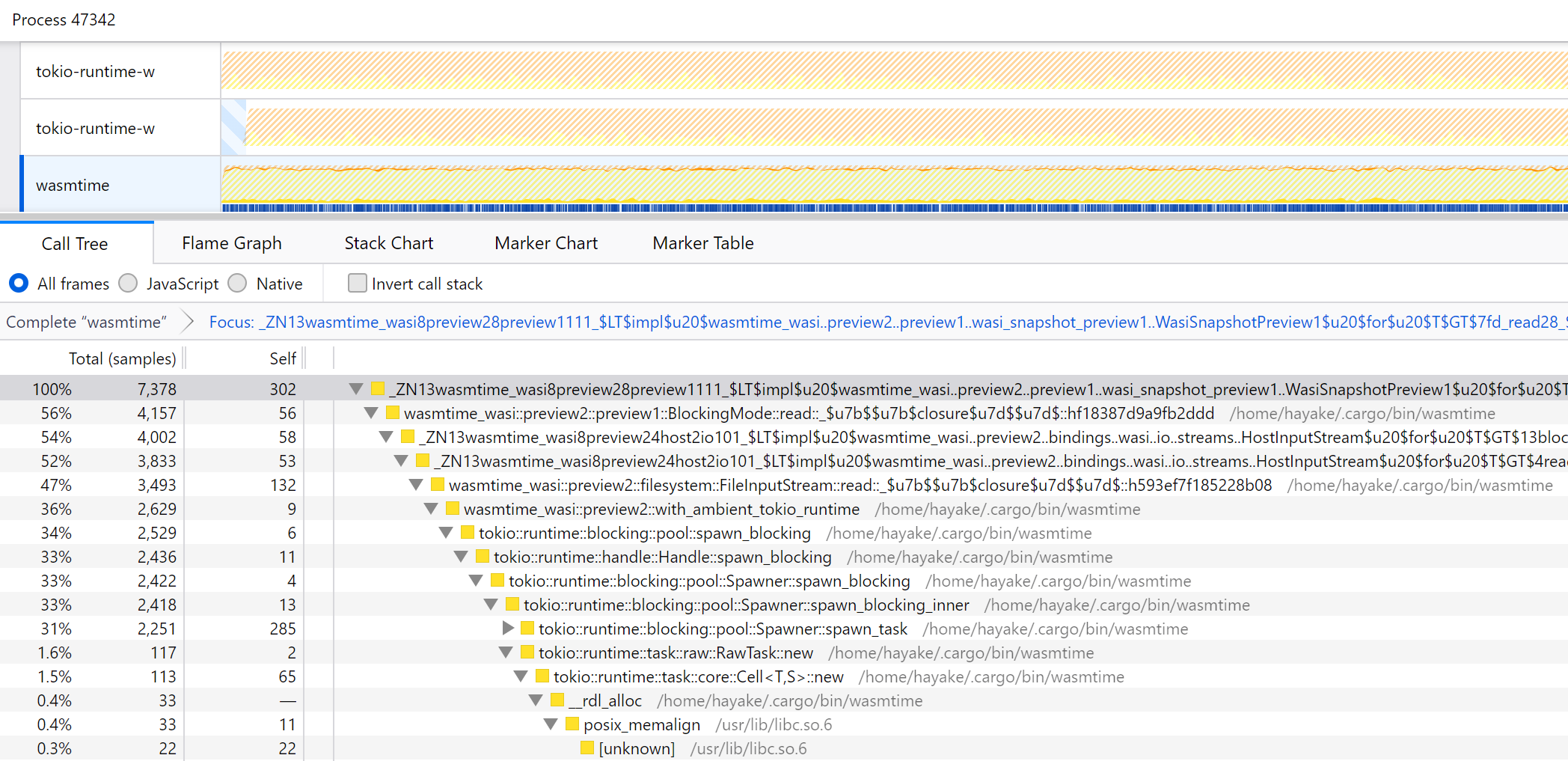}
    \caption{Wasmtime\_preview2 wasmtime process executes read WASI}
    \label{fig:tokioread}
\end{figure}
\begin{figure}[!htb]
    \centering
    \includegraphics[width=0.45\textwidth]{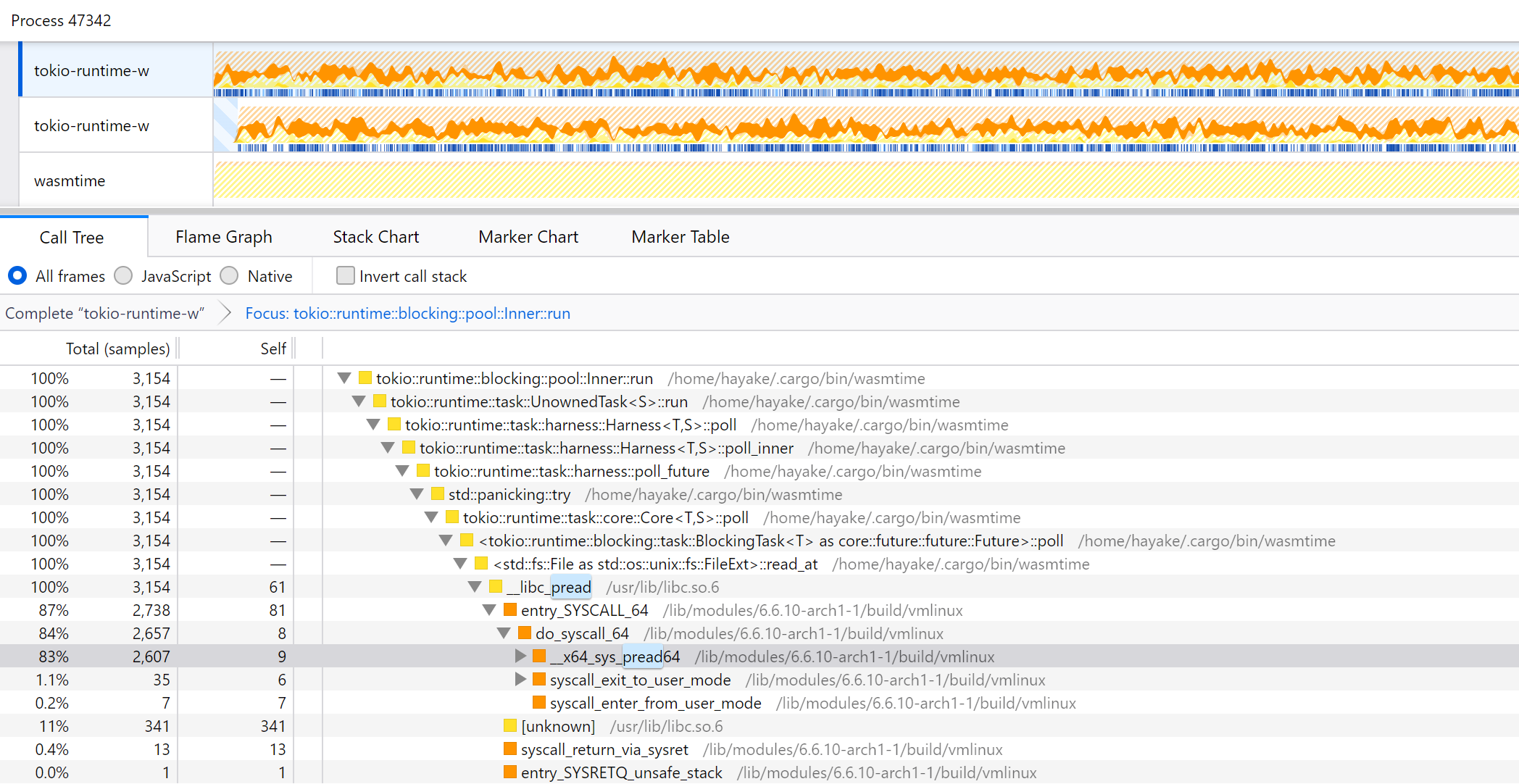}
    \caption{Wasmtime\_preview2 tokio-runtime-w process asynchronously executes \_x64\_sys\_pread}
    \label{fig:tokiopread}
\end{figure}

Case 2 analyzes the impact of the asynchronous library on two runtimes, Wasmer and Wasmtime\_preview2. As shown in Figure\ref{fig:tokioread} and Figure \ref{fig:tokiopread}. 

Wasmtime\_preview2 uses the Tokio asynchronous library. When performing I/O operations, three processes are executed in total. 
The runtime's main process handles the \textit{read} WASI calls (symbol \textit{\_ZN13wasmtime\_wasi8preview28...}), while Tokio starts two asynchronous processes, with one of these processes executing the \textit{pread} syscall (symbol \textit{\_x64\_sys\_pread64} within \textit{\_libc\_pread}), which is asynchronous with the runtime's own WASI process. Similarly, Wasmer also uses this asynchronous library. Under heavy I/O load, these two runtimes exhibit certain peculiarities in their WASI and syscalls compared to other runtimes.

In Figure \ref{fig:rq1syscall}, the \textit{write} syscall time for Wasmer and Wasmtime\_preview2 is significantly longer than other runtimes. 
Referring to Table \ref{tab:fwrite}, it can be observed that the syscall times for several runtimes are quite similar, around 2100 ns. 
However, the number of syscalls for Wasmer and Wasmtime\_preview2 is twice that of the other runtimes.

\begin{figure}[!htb]
    \centering
    \includegraphics[width=0.45\textwidth]{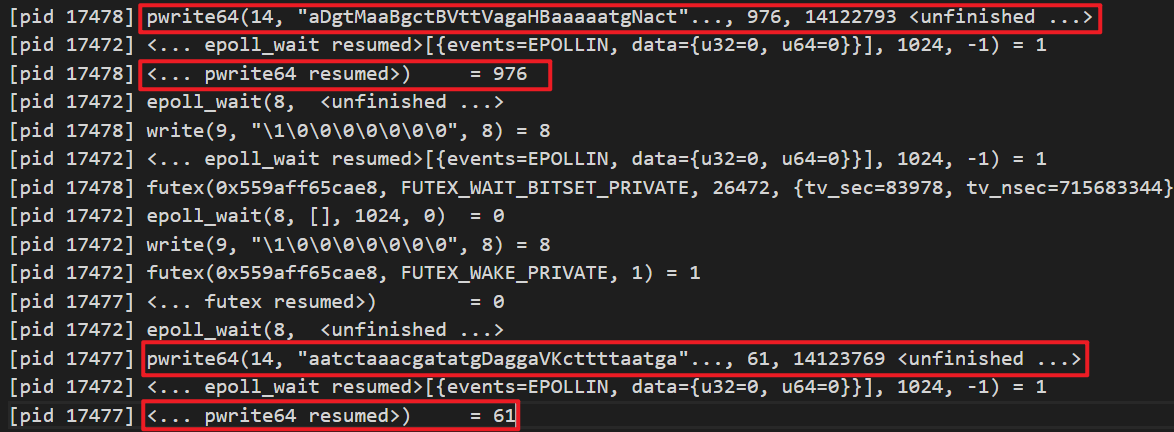}
    \caption{Wasmtime\_preview2 Syscall: pwrite64}
    \label{fig:3}
\end{figure}

\begin{figure}[!htb]
    \centering
    \includegraphics[width=0.45\textwidth]{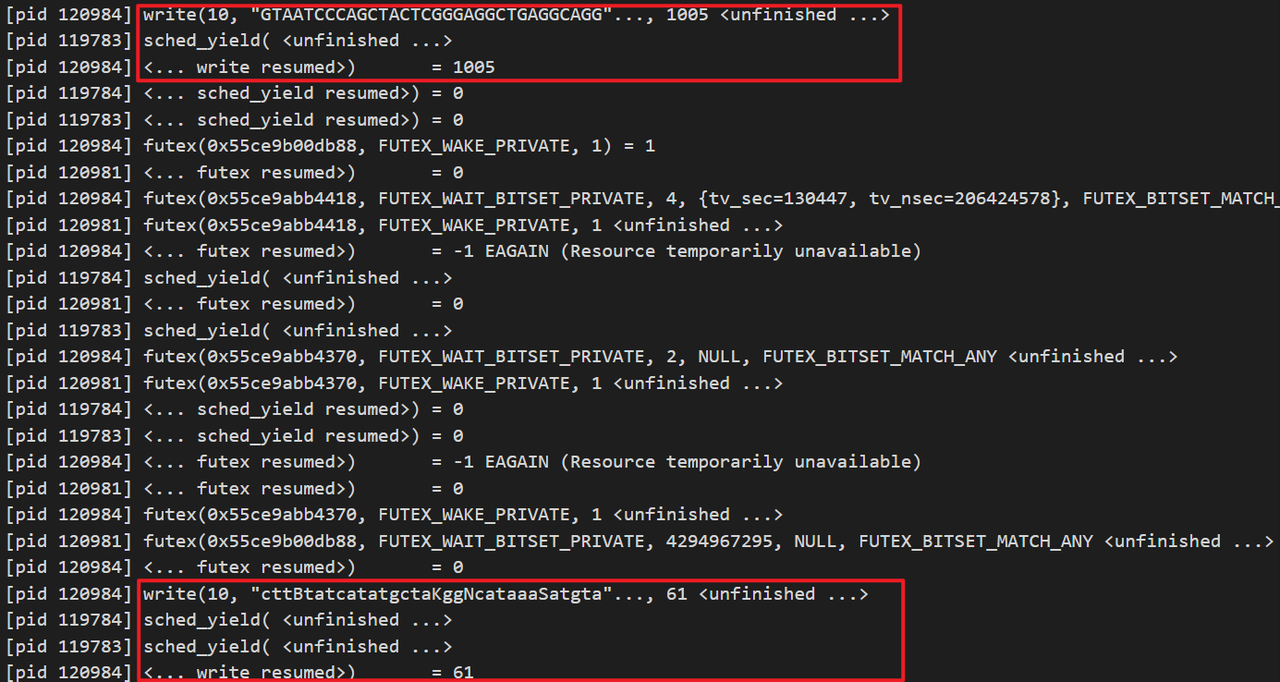}
    \caption{Wasmer Syscall: write}
    \label{fig:4}
\end{figure}

Using the Strace tool, the syscall processes for the two runtimes are traced. 
Figure \ref{fig:3} shows the syscalls for Wasmtime\_preview2, where other runtimes read two iov vectors in a single \textit{writev} call, whereas Wasmtime\_preview2 uses two separate \textit{pwrite64} syscalls. 
Similarly, Figure \ref{fig:4} illustrates the syscalls of Wasmer. It also uses two separate \textit{write} syscalls for reading. 
Given that the time for each syscall does not differ significantly, the doubled number of syscalls results in a longer total syscall time.


Tokio chooses to break down large \textit{writev} into multiple smaller \textit{write} due to the characteristics of the asynchronous library, considering the time-slice rotation mechanism of the operating system. 
A larger \textit{writev} call might result in a longer execution time, delaying the execution of other tasks. 
Breaking it down into smaller operations can reduce long-time-slice occupancy, avoiding I/O or CPU bottlenecks caused by a single large operation. 
Additionally, breaking down operations increases the likelihood of completing them within a single time slice, thereby reducing the need for the operating system to perform context switches between tasks. 
In an asynchronous environment, executing short operations allows the asynchronous library to respond more quickly to other tasks, which is crucial for building high-performance asynchronous systems.

As shown in Figure \ref{fig:rq1wasi}, the WASI execution time for Wasmer and Wasmtime\_preview2 is significantly longer than that for other runtimes. 
From Section \ref{rq2}, it can also be seen that the WASI execution time for these two runtimes accounts for a larger proportion of the total runtime.


This is because the Tokio asynchronous library used in Wasmtime\_preview2 has not been specifically optimized for non-asynchronous scenarios. 
Tokio uses \textit{spawn\_blocking} to implement blocking. 
Based on the configuration of the \textit{allow\_blocking\_current\_thread} variable within the object, it decides whether to synchronously execute a closure in the current execution environment or asynchronously execute it through Tokio's blocking thread pool to avoid blocking the current asynchronous execution environment.

Although \textit{spawn\_blocking} in Tokio is designed to handle synchronous operations that may cause blocking in an asynchronous execution environment by running these operations on dedicated threads, it aims to keep the asynchronous executor running efficiently rather than being blocked by a time-consuming synchronous operation. 
However, the test cases used in this paper are synchronously sequential programs, and remains in a blocked-wait state during I/O operations without executing other computational tasks to improve performance. 
As shown in Figure \ref{fig:3} and Figure \ref{fig:4}, this instead leads to additional context switching, synchronization mechanisms (futex), and thread management overhead, thereby introducing unnecessary performance burdens and adversely affecting WASI's interaction performance during I/O operations.

In the absence of sufficient parallel tasks or asynchronous logic to fully utilize the advantages of multithreading, these overheads may not be offset by the blocking time saved. In fact, in such cases, directly executing I/O synchronously may be more efficient because it avoids the complexity and additional resource consumption introduced by multithreading. Recently, to optimize this process \cite{tokioperformance}, the open-source community has provided a flag option for \textit{wasiCtx} that allows users to choose whether to call \textit{spawn\_blocking} to execute on the current thread, thus avoiding blocking the main thread in synchronous scenarios.

\section{RELATED WORK}

\subsection{Wasm Reliability}
Some studies have empirically explored bugs related to Wasm \cite{hilbig2021empirical,wang2023comprehensive,romano2021empirical}. Hilbig et al. \cite{hilbig2021empirical} focused on bugs within Wasm modules, collecting and analyzing Wasm binaries from various real-world sources. 
Wang et al. \cite{wang2023comprehensive} conducted the first empirical analysis of 867 bugs in four popular Wasm runtimes, uncovering the characteristics, impacts, and repair patterns of these bugs. 
They discussed the broad implications of their findings for bug detection, localization, debugging, and repair in Wasm runtimes. 
Romano et al. \cite{romano2021empirical} performed qualitative and quantitative analyses of 1,200 bugs in three open-source Wasm compilers, deeply investigating the characteristics and impacts of these bugs, providing insights for designing tools to test and debug Wasm compilers. 

Other research has aimed to explore methods to ensure the security of Wasm module execution, whether by implementing sandboxing to ensure system security \cite{bosamiya2022provably,abbadini2023poster} or by discovering inconsistencies in Wasm module execution \cite{zhou2023wadiff,cao2023wrtester}.  
Abbadini et al. \cite{abbadini2023poster} proposed a method to enhance Wasm runtime sandboxing by using eBPF programs instead of traditional security checks, achieving finer-grained access security restrictions for the file system. Zhou et al. \cite{zhou2023wadiff} designed a symbolic execution engine to generate test cases that can trigger runtime inconsistent behavior, proposing the differential testing framework WADIFF to evaluate the accuracy and reliability of Wasm runtimes. 
Cao et al. \cite{cao2023wrtester} proposed the differential testing framework WRTester, which generates complex Wasm test cases by decomposing and assembling real-world Wasm binaries to trigger hidden inconsistencies between independent runtimes. 
They designed a runtime-independent root cause localization method to accurately pinpoint errors in independent runtimes.

\subsection{Wasm Performance Testing}
Unlike traditional JS, Wasm as an emerging compilation target provides performance closer to native execution in the browser, empowering the development of web applications. 
In recent years, significant efforts have been focused on comparing various metrics of performance between JS and Wasm in the browser \cite{yan2021understanding,wang2021empowering,de2021runtime}, providing directions for future optimizations of Wasm technology. 
The work by Yan et al. \cite{yan2021understanding} and Wang \cite{wang2021empowering} systematically reveals the characteristics and differences between Wasm and JS in terms of memory usage, execution time, and performance optimization. They also investigate whether browsers can optimize Wasm execution compared to JS. Macedo et al. \cite{de2021runtime} construct a benchmarking framework to explore whether Wasm has advantages in energy efficiency and runtime compared to JS. Romano et al. \cite{romano2023function} elucidate how browser compilers optimize the execution efficiency of Wasm modules and study the counter-intuitive effects of function inlining optimization on the performance of Wasm module execution.

Regarding performance testing of server-side Wasm, there is relatively less research \cite{spies2021evaluation,jiang2023revealing}. Spies et al. \cite{spies2021evaluation} measure metrics such as code size, startup time, and execution time in non-web environments, comparing different Wasm implementations with native code and JS optimized using asm.js. 
Jiang et al. \cite{jiang2023revealing} design a differential testing method called WarpDiff to identify performance issues in server-side Wasm runtimes and further analyze exceptional cases, summarizing seven popular categories of performance issues in independent runtime implementations of Wasm, sparking research into improving Wasm independent runtime implementations.

However, existing research primarily focuses on identifying unreasonable implementations based on the overall execution time and final state of Wasm modules, with a lack of studies on testing the efficiency of independent runtime WASI implementations, making it difficult to evaluate the performance of runtime I/O interactions with the operating system via WASI.

\section{FUTURE WORK}
In this work, we currently focus only on empirical analysis of I/O performance data, revealing performance issues in WASI during I/O processes in independent Wasm runtimes. 
By analyzing the causes of performance anomalies in two specific cases, the method provides recommendations for future performance improvements in runtime WASI. However, in this paper, we do not yet propose specific improvements for the performance issues in WASI implementations. 
Future research needs to include experiments to refine performance improvements.

Besides, the proposed method automatically obtain performance data for WASI, but the identification of performance anomalies and the analysis of their causes still require manual intervention, which is inefficient. In the future, it is hoped that methods from other fields for automatic anomaly detection can be referenced to achieve automated identification of performance anomaly data and specific code locations.

Additionally, since eBPF tools and the BCC framework are only applicable on Linux systems and certain system events on Windows are relatively closed, the tools developed in this paper currently only support performance monitoring on Linux systems. Future research will explore suitable methods to capture syscall events on Windows, making the tools applicable to a wider range of systems.

\section{CONCLUSION}

With the development of independent Wasm runtimes, Wasm is no longer confined to the browser but also offers a fast, scalable, secure, and sandboxed way to run the same code across all machines on the server side. 
In contrast to browser environments, the design proposals and implementations of the WebAssembly System Interface (WASI) remain in their nascent stages. 
Moreover, the implementation of the WASI interface exhibits considerable variability across different server-side runtimes.
Inefficient interface implementations can slow down runtime interactions with the system, making it difficult to highlight Wasm’s performance advantages. 
Addressing this issue, in this paper, we proposes an eBPF-based performance analysis method for Wasm runtime I/O operations. 
By collecting performance data, the proposed framework aims to identify performance issues in the WASI implementations within Wasm runtimes. 
The framework does not only provide strong technical support for improving the performance of Wasm module execution in independent runtimes from the perspective of WASI implementation mechanisms but also stimulates future optimization efforts for more efficient independent runtimes and WASI implementations.

\section{Acknowledgments}
The work described in this paper was supported by the National Natural Science Foundation of China (No. 62202511) and Guangdong Natural Science Foundation General Program (Grant No. 2022A1515011713).

\bibliographystyle{IEEEtran}
\bibliography{bibref.bib}

\end{document}